\documentstyle[pra,aps,multicol,array]{revtex}
\newcommand{\beq}{\begin{equation}}
\newcommand{\eeq}{\end{equation}}
\newcommand{\beqa}{\begin{eqnarray}}
\newcommand{\eeqan}{\end{eqnarray*}}
\newcommand{\beqan}{\begin{eqnarray*}}
\newcommand{\eeqa}{\end{eqnarray}}
\newcommand{\bra}[1]{\left\langle{#1}\right|}
\newcommand{\ket}[1]{\left|{#1}\right\rangle}
\newcommand{\ip}[1]{\left\langle{#1}\right\rangle}
\newcommand{\non}{\nonumber}
\newcommand{\id}{\openone}
\newcommand{\eq}[1]{Eq.~(#1)}
\newcommand{\eqr}[1]{Eq.~(\ref{#1})}

\begin{document}
\title{A class of symmetric controlled quantum operations}
\author{John A. Vaccaro, O. Steuernagel and S.F. Huelga}
\address{Division of Physics and Astronomy. Department of Physical Sciences,\\
University of Hertfordshire, Hatfield AL10 9AB, England}
\date{\today}
\maketitle

\begin{abstract}
Certain quantum gates, such as the controlled-NOT gate, are
symmetric in terms of the operation of the control system upon
the target system and vice versa.  However, no operational
criteria yet exist for establishing whether or not a given
quantum gate is symmetrical in this sense.  We consider a
restricted, yet broad, class of two-party controlled gate
operations for which the gate transforms a reference state of the
target into one of an orthogonal set of states.  We show that for
this class of gates it is possible to establish a simple
necessary and sufficient condition for the gate operation to be
symmetric.
\end{abstract}


\section{Introduction}

Quantum gates are the building blocks for quantum information
hardware \cite{gates}. In particular, non-local quantum gates
involving remote processors are the essential ingredients for
performing distributed quantum computation \cite{distri} and
implementing quantum communication protocols \cite{innsbruck}. A
given gate operation can be characterised in terms of how much
entanglement it can create and how much classical information it
can convey. This parameterisation establishes lower bounds in both
the amount of entanglement and the classical communication
resources that are required for the optimal implementation of a
given quantum gate \cite{tony,poly,popescu,cirac}.

In this paper, we consider a particular class of controlled gates
whose action is to transform a reference state of the target into
one of a set of $N$ orthogonal states.  For clarity we shall call
these gates {\em orthogonal gates}. They are generalisations of
the controlled-NOT (CNOT) gate. Our aim is to provide an
operational criteria which ensures {\em symmetric} operation of an
orthogonal gate in the sense that the {\em control can be swapped
with the target} for suitable preparation of the input states.
Symmetric orthogonal gates allow the control system to communicate
$\log_2N$ bits of classical information to the target system and,
because of their symmetry, allow for the same amount of reverse
classical communication from target to control. By a suitable
change of basis, they can also generate mutual entanglement
between the states of the control and target systems. The
orthogonality property ensures that an entanglement of $\log_2N$
ebits can be generated in this way. Since there is a simple
connection between the classical communication and the
entanglement generated for symmetric orthogonal gates, we focus
only on the classical information capability between two parties
in this paper.

The main ideas underlying symmetric gates can be illustrated by
considering a couple of specific gates.  The CNOT gate is the
simplest quantum gate. It is also a prime example of a symmetric
gate:  if Alice has the control qubit and Bob the target qubit,
Alice can transmit one classical bit of information to Bob using
the computational basis, and conversely Bob can transmit one
classical bit of information to Alice using a Hadamard
transformation of the computational basis \cite{poly,popescu}.

The gate at the next level of sophistication is the
controlled-Pauli (CP) gate \cite{mbp}.  This gate applies either
the identity or one of the three Pauli-operators $\{\sigma_i:
i=x,y,z\}$ on a target qubit depending on the state of two
control qubits, and can be written as
\beqan
    U_{CP} &=& \ket{00}\bra{00} \otimes \id
                   + \ket{01}\bra{01} \otimes \sigma_x\\
           &+& \ket{10}\bra{10} \otimes \sigma_y
                    + \ket{11}\bra{11} \otimes \sigma_z .
\eeqan
Let Alice hold the control qubits and Bob the target qubit. Alice
can transmit 2 classical bits of information per application of
the CP gate using a variation of super-dense coding
\cite{superdense}. To see this consider the following protocol.
Alice encodes a 2-bit message in binary notation using the basis
states $|00\rangle,|01\rangle,|10\rangle,|11\rangle$ of her two
qubits. Assume that Bob's qubit is in the (entangled) Bell state
$\ket{\phi^+}=\ket{00}+\ket{11}$ with another qubit at his site.
The CP gate is applied between Alice's particle and the first of
Bob's particles. Depending on the state in which Alice has
prepared her two control qubits, Bob will subsequently hold one
of the four Bell states, which are mutually orthogonal.
Therefore, he is able to infer Alice's message and thus receive 2
classical bits of information from Alice. We would like to know
whether the gate is symmetrical from the point of view of its
classical information capacity. In other words, {\it is it
possible for Bob to choose certain initial states so that the
gate operation results in two classical bits being conveyed to
Alice?} If so, we would say that the gate is symmetric.

We can easily show that the CP gate can transmit one classical
bit from Bob to Alice as follows. Consider the case when the
first of Alice's qubits is kept in a fixed state, for instance in
state $\ket{0}$. The action of a CP gate is now equivalent to a
controlled-NOT gate between Alice's second qubit and Bob's first
qubit. Given that a controlled-NOT gate can transmit one
classical bit in each direction \cite{poly,popescu}, we know
immediately that Bob can convey at least one classical bit to
Alice. However we cannot be sure that certain initial preparation
may allow Bob to actually transmit two classical bits using the
CP gate. The aim of the paper is to remove this ambiguity for
general orthogonal gates by providing an operational criteria for
establishing the symmetry of this class of controlled gates.

The organisation of the remainder of the paper is as follows. We
define orthogonal gates and present the conditions for symmetric
orthogonal gates in Section \ref{theorem}, we then apply the
conditions to a number of different gates in Section
\ref{applications}, we give proofs of the conditions in Section
\ref{proof} and we end with a discussion in Section
\ref{discussion}.

\section{Symmetric orthogonal gates} \label{theorem}

A controlled quantum gate $\cal G$ allows Alice, using different
orthogonal quantum states $\ket{\psi_n}_A$, to control the
application of a set of unitary operations $\{ U_n
\}$ on Bob's particle's state $\ket{\phi}_B$. We call the gate's
operation a {\it symmetric} controlled operation if it allows Bob
to conversely control the state of Alice's particle to the same
degree. As mentioned above, the CNOT is an example of a symmetric
controlled gate with the controlled operations $\{U_n\}$ being
the identity and the $\sigma_x$ Pauli operator.

We restrict our attention here to the class of quantum gates for
which the $N$ controlled unitary operations $\{U_n\}$ operating on
Bob's state space produce a set of $N$ orthogonal states
\beq
  {\cal N} =\{\ket{n}_B: {}_B\ip{n|m}_B=\delta_{n,m};\ n,m=1,...,N\}
  \label{setN}
\eeq
from a fixed reference state $|R\rangle_B$:
\beq
  U_n|R\rangle_B = |n\rangle_B \mbox{ for } n=1,\ldots,N
  \label{ortho1}
\eeq
and so ${}_B\!\ip{R|U^\dagger_m U_n|R}_B=\delta_{m,n}$.  We
shall call such gates {\em orthogonal gates of cardinality $N$}.
For clarity, we use subscripts $A$ and $B$ here, and subsequently,
to distinguish the states of Alice's and Bob's particles,
respectively. Note that we do not assume that $\ket{R}_B \in \cal
N$ nor that ${\bf\id} \in \{U_n\}$. Clearly the dimension of the
state spaces of Alice's and Bob's particles must be at least $N$.
For this paper we assume that the dimension of Bob's state space
is {\em exactly} $N$. The (unitary) action of the gate $\cal G$ is
hence assumed to be of the form
\beqa
  {\cal G} \left( \ket{\psi}_A \ket{\phi}_B\right)
     &=& \sum_{n=1}^N a_n \ket{n}_A U_n\ket{\phi}_B
                 \label{input1} \\
     &=& \sum_{n,m=1}^N a_n \ket{n}_A b_m \; U_n \ket{m}_B
                 \label{2gate}
\eeqa
where the input states are given by $\ket{\psi}_A=\sum_{n} a_n
\ket{n}_A$ and $\ket{\phi}_B=\sum_{n} b_n \ket{n}_B$ with
$\sum_n|a_n|^2=\sum_n|b_n|^2=1$.  Using Eqs.~({\ref{ortho1}}) and
(\ref{2gate}) it is easy to show that Alice can send one of $N$
distinct messages to Bob in each application of the gate when the
following input states are employed:
\beqan
    \ket{\psi_n}_A=\ket{n}_A, \; n=1,...,N \; \;  \mbox{  and  }
    \ket{\phi}_B= \ket{R}_B \ .
\eeqan

To be a symmetric gate, $\cal G$ must allow Bob to send one of
$N$ distinct messages to Alice per application of the gate for
some suitable choice of input states. We prove the following
equivalent Theorems in Section \ref{proof}.

{\bf Theorem 1.} A necessary and sufficient condition for
symmetric operation of an orthogonal gate of cardinality $N$ is
that the set of pairwise products $\{U^\dagger_n U_m:
n,m=1,\ldots,N\}$ of the controlled operations $U_n$ have the
commuting property
\beq
  (U^\dagger_n U_m) (U^\dagger_p U_q) = (U^\dagger_p U_q) (U^\dagger_n
  U_m) \; ,
  \label{commute}
\eeq
for $n,m,p,q=1,...,N$.

{\bf Theorem 1$'$.} A necessary and sufficient condition for
symmetric operation of an orthogonal gate of cardinality $N$ is
that the set of controlled operators $\{U_n\}$ is related to a set
of $N$ commuting operators $\{C_n\}$ by
\beq
  U_n=T C_n \mbox{ where }
    C_n C_m = C_m C_n \mbox{ for } n,m=1,\ldots,N
   \label{equiv_commute}
\eeq
and where $T$ is unitary.

The latter version of the theorem is perhaps more transparent but,
if $T$ is not known, it is generally more straightforward to check
the commuting property using \eq{\ref{commute}}.  We note that if
the set $\{U_n\}$ includes the identity then \eqr{commute} implies
the operators $U_n$ must be mutually commuting.  We shall also
prove the following Theorem and Corollary regarding the structure
of the states involved.

{\bf Theorem 2.}  Bob is able to send one of $N$ distinct
messages to Alice using a symmetric orthogonal gate of
cardinality $N$ iff the input state of Bob's particle is an
eigenstate of the pairwise product $U_n^\dagger U_m$ [or,
equivalently, an eigenstate of the operators $C_n$ in
\eqr{equiv_commute}], and Alice's input state is
$\case{1}{\sqrt{N}}\sum_n e^{i\eta_n}\ket{n}_A$ for arbitrary,
real $\eta_n$.

{\bf Corollary 3.} The reference state $\ket{R}_B$ and basis set
${\cal N}$ in \eqr{ortho1} for a symmetric orthogonal gate of
cardinality $N$ are given by
\beqa
   \ket{R}_B &=& \frac{1}{\sqrt{N}}\sum_{r=1}^N
              e^{-i\gamma_r}\ket{\lambda_r}_B    \label{Rstate}\ , \\
   \ket{n}_B &=& \frac{1}{\sqrt{N}}\sum_{r=1}^N
              e^{i[\varphi_n(r)-\gamma_r]}U_1\ket{\lambda_r}_B
              \mbox{ for }n=1,\ldots,N   \label{exp_setN}
\eeqa
where $ e^{i\varphi_n(r)}$ and $\ket{\lambda_r}_B$ are the $r$th
eigenvalue and eigenstate of the product $U_1^\dagger U_n$, and
$\gamma_r$ is an arbitrary, real parameter.  Conversely, the
eigenstates are given in terms of the ${\cal N}$-basis as
\beq
   \ket{\lambda_r}_B
       =U_1^\dagger\frac{1}{\sqrt{N}}e^{i\gamma_r}
             \sum_{n=1}^N e^{-i\varphi_n(r)}\ket{n}_B
             \mbox{ for }r=1,\ldots,N\ .
             \label{lambda}
\eeq

Whereas Theorems 1 and $1'$ are useful for deciding whether a
particular orthogonal is symmetric or not, Theorem 2 and Corollary
3 are useful for constructing a symmetric orthogonal gate from a
set of operators satisfying Eqs. (\ref{commute}) and
(\ref{equiv_commute}).

\section{Applications}  \label{applications}

We illustrate the application of the Theorems and Corollary with a
few examples. The simplest example is given for cardinality $N=2$,
where $N$ is the size of the set of controlled operators
$\{U_n\}$. This occurs for the controlled-U gate where $\{U_n\}$
contains the identity $U_1=\id$ and another operator $U_2$. It is
straightforward to show that this gate satisfies the orthogonal
property Eq.~(\ref{ortho1}) when ${\cal N}$ is the computational
basis iff $U_2$ anticommutes with the operator $\sigma_z$. That
is, $U_2$ is of the form \beqan U_2 = e^{i \alpha} \, \left(
    \begin{array}{cc}
    0 & b\\
    -b^* &  0
    \end{array}
    \right)
\eeqan
where $\alpha$ is real and $|b|^2=1$. The controlled-NOT, with
$\alpha=0$ and $b=1$, is an example of this class. Since the set
$\{U_n\}$ contains only two operators, one of which is the
identity, the condition \eqr{commute} is clearly satisfied. Hence
all orthogonal controlled-U gates of cardinality $2$ are
symmetric.

Next consider the controlled-Pauli gate.  In this case the set
$\{U_n\}$ of controlled operators is given by
$\{\id_1\otimes\id_2,\sigma_x\otimes\id_2,\sigma_y\otimes\id_2,
\sigma_z\otimes\id_2\}$ and so the cardinality is $N=4$. Here
$\id_i$ is the identity operator associated with Bob's $i$th
qubit.  This set fulfills the orthogonal property \eqr{ortho1}
when acting on the Bell state $\ket{\phi^+}_B$, as discussed in
the Introduction, and so it is an orthogonal gate. But it fails to
fulfill the condition \eqr{commute} for all values of the indexes
$m,n,p,q$ and so the controlled Pauli gate is not a symmetric
gate. This means that if Alice has the control, Bob {\em cannot}
send 2 classical bits to Alice in one application of the gate. In
the final Section, we determine the maximum amount of information
Bob can actually send using this gate.

There are, however, symmetric orthogonal gates with cardinality
$N=4$. One is given by the following set of commuting operators:
\beqan
  C'_1 &=& \ket{\lambda_1}_{B\ B}\!\bra{\lambda_1}
          +\ket{\lambda_2}_{B\ B}\!\bra{\lambda_2}
          +\ket{\lambda_3}_{B\ B}\!\bra{\lambda_3}
          +\ket{\lambda_4}_{B\ B}\!\bra{\lambda_4}  \\
  C'_2 &=& \ket{\lambda_1}_{B\ B}\!\bra{\lambda_1}
          -\ket{\lambda_2}_{B\ B}\!\bra{\lambda_2}
          +\ket{\lambda_3}_{B\ B}\!\bra{\lambda_3}
          -\ket{\lambda_4}_{B\ B}\!\bra{\lambda_4}  \\
  C'_3 &=& \ket{\lambda_1}_{B\ B}\!\bra{\lambda_1}
          +\ket{\lambda_2}_{B\ B}\!\bra{\lambda_2}
          -\ket{\lambda_3}_{B\ B}\!\bra{\lambda_3}
          -\ket{\lambda_4}_{B\ B}\!\bra{\lambda_4}  \\
  C'_4 &=& \ket{\lambda_1}_{B\ B}\!\bra{\lambda_1}
          -\ket{\lambda_2}_{B\ B}\!\bra{\lambda_2}
          -\ket{\lambda_3}_{B\ B}\!\bra{\lambda_3}
          +\ket{\lambda_4}_{B\ B}\!\bra{\lambda_4}  \\
\eeqan
for an arbitrary basis $\{\ket{\lambda_n}\}$.  Eqs. (\ref{setN})
and (\ref{ortho1}) are satisfied for a suitable reference state,
such as $\ket{R}_B=\case{1}{2}\sum_n\ket{\lambda_n}_B$,
constructed using \eqr{Rstate}. Also, one can easily show using
\eqr{input1} that for the input state $\ket{\psi}_A
\ket{\phi_r}_B=\case{1}{2}\sum_n\ket{n}_A
\ket{\lambda_r}_B$ this gate produces one of four
possible output states as follows
\beqan
 {\cal G} \left( \ket{\psi}_A \ket{\phi}_B\right) = \left\{
   \begin{array}{ll}
    \case{1}{2}(\ket{1}_A+\ket{2}_A+\ket{3}_A+\ket{4}_A)\ket{1}_B & \mbox{for } r=1 \\
    \case{1}{2}(\ket{1}_A-\ket{2}_A+\ket{3}_A-\ket{4}_A)\ket{2}_B & \mbox{for } r=2 \\
    \case{1}{2}(\ket{1}_A+\ket{2}_A-\ket{3}_A-\ket{4}_A)\ket{3}_B & \mbox{for } r=3 \\
    \case{1}{2}(\ket{1}_A-\ket{2}_A-\ket{3}_A+\ket{4}_A)\ket{4}_B & \mbox{for } r=4
  \end{array}
  \right.
\eeqan Alice can distinguish between the four possible final
states of her particle because they are orthogonal, and so Bob can
send 2 classical bits of information, per application of the gate,
to Alice by his choice of input state $\ket{\lambda_r}_B$.  The
gate is therefore symmetric.

The simplest example of a symmetric orthogonal gate for arbitrary
cardinality $N$ is given by what we call the {\em
controlled-shift gate} where the operation on the target is one
of the family of $N$ shift operators $C''_{n}$. The shift
operators are defined as $C''_{n} \equiv \sum_{m=1}^N\ket{(n-2+m)
\mbox{mod}N+1}_{B\ B}\!\bra{m}$ for $n=1,\ldots,N$.
These operators produce a cyclic shift amongst the basis states
${\cal N}$ and clearly satisfy the orthogonal property
\eqr{ortho1} for arbitrary reference state $\ket{R}_B\in {\cal
N}$. It is easy to show that the set of shift operators
$\{C''_n\}$ satisfies the commuting condition \eqr{equiv_commute}
with $T=\id$ and so the controlled-shift gates are clearly
symmetric orthogonal gates. Similarly, the related controlled-U
gate which applies the unitary operation $T C''_n$, for some
nontrivial unitary $T$, to the target particle is also a
symmetric orthogonal gate.

\section{Proofs} \label{proof}

In this Section, we first prove the equivalence of the two
Theorems $1$ and $1'$.  We then give the proof of these Theorems
by showing the sufficiency and necessity of the criteria
\eqr{commute} for symmetric operation.  We end with proofs of
Theorem 2 and Corollary 3.

{\bf Proof of equivalence.}  The commuting property \eqr{commute}
implies that the (unitary) products $U^\dagger_n U_m$ share the
same set of $N$ eigenvectors $\{\ket{\lambda_n}_B\}$. Consider
the eigenvalue equation for the products $U^\dagger_1 U_m$ for
$m=1,\ldots,N$ on the $r$th eigenvector \cite{note_U_1}:
\beq
  U^\dagger_1 U_m \ket{\lambda_r}_B = e^{i\varphi_m(r)}
  \ket{\lambda_r}_B\ ,
  \label{eigen}
\eeq
where the $\varphi_m(r)$ are real parameters with
$\varphi_1(r)=0$; rearranging gives
\beq
  U_m \ket{\lambda_r}_B = e^{i\varphi_m(r)} U_1\ket{\lambda_r}_B
  \label{diag}
\eeq
and hence
\beq
   U_m = U_1 \sum_{r=1}^N e^{i\varphi_m(r)}
   \ket{\lambda_r}_{B\ B}\!\bra{\lambda_r} \ .
   \label{Um}
\eeq
since the states $\{\ket{\lambda_r}_B\}$ form a complete
orthonormal basis. We write $T=U_1\sum_n
e^{-iw_r}\ket{\lambda_r}_{B\ B}\!\bra{\lambda_r}$ for arbitrary
real $w_r$.  Then
\[
   U_1 = T \sum_{r=1}^N e^{iw_r}\ket{\lambda_r}_{B\ B}\!\bra{\lambda_r}
\]
and so from \eqr{Um} we now have $U_m=T C_m$ where the $C_m$,
\[
   C_m =
   \sum_{r=1}^N e^{i[\varphi_m(r)+w_r]}\ket{\lambda_r}_{B\
   B}\!\bra{\lambda_r}\ ,
\]
form a set of $N$ mutually-commuting operators. This shows that
\eqr{equiv_commute} follows from \eqr{commute}.  The converse,
that \eqr{commute} follows from \eqr{equiv_commute}, is trivially
true.  This completes the proof of the equivalence of the two
Theorems 1 and $1'$.

{\bf Proof of Theorems 1 and 1$'$.} The {\em sufficiency} of the
commuting condition \eqr{commute} is proved by showing that it
allows Bob to generate one of $N$ orthogonal output states of
Alice's particle depending on the input state of his particle.
Consider the trace of the product $U_n^\dagger U_m$ in the ${\cal
N}$-basis \eqr{setN}:
\[
  {\rm tr}(U_n^\dagger U_m) = {\rm tr}(U_m U_n^\dagger)
  = \sum_{r=1}^N {}_B\!\bra{r}U_m U_n^\dagger\ket{r}_B \ .
\]
Using ${}_B\!\bra{r}U_m
U^\dagger_n\ket{r}_B={}_B\!\bra{R}U^\dagger_r U_m U^\dagger_n
U_r\ket{R}_B$ from \eqr{ortho1}, $U^\dagger_r U_m U^\dagger_n
U_r=U^\dagger_n U_r U^\dagger_r U_m$ from \eqr{commute} and the
unitarity property $U_r U^\dagger_r=\id$ gives \beqa
  {\rm tr}(U_n^\dagger U_m)
            = \sum_{r=1}^N {}_B\!\ip{R|U^\dagger_n U_m|R}_B
            = \sum_{r=1}^N  {}_B\!\ip{n|m}_B
            = N\delta_{n,m}
  \ .  \label{tr1}
\eeqa
However, calculating the trace in the eigenbasis
$\{\ket{\lambda_r}_B\}$ given by \eqr{eigen} is found to give
\beqa
  {\rm tr}(U_n^\dagger U_m)
    &=& \sum_{r=1}^N {}_B\!\bra{\lambda_r} U_n^\dagger U_m \ket{\lambda_r}_B\non\\
    &=& \sum_{r=1}^N e^{i[\varphi_m(r)-\varphi_n(r)]}
                            \label{tr2}
\eeqa
on use of \eqr{diag}.  Equating \eqr{tr1} and \eqr{tr2} then
yields
\beqa
 \delta_{n,m} = \frac{1}{N}\sum_{r=1}^N
                e^{i[\varphi_m(r)-\varphi_n(r)]}\ .
  \label{ortho_eigen1}
\eeqa
Consider the $N$-dimensional vector ${\bf v}(n)$ whose elements
are given by $v_r(n)= e^{i\phi_n(r)}/\sqrt{N}$ for
$r=1,\ldots,N$.  \eqr{ortho_eigen1} shows that the set of vectors
${\bf v}(n)$ for $n=1,\ldots,N$ form an orthonormal set, and so
the matrix ${\bf M}$ whose columns are given by ${\bf v}(n)$ is
unitary: ${\bf M}^\dagger{\bf M}={\bf M M}^\dagger={\bf\id}$. The
expression ${\bf M M}^\dagger={\bf\id}$ implies that
\beqa
 \delta_{r,r'} = \frac{1}{N}\sum_{n=1}^N e^{i[\varphi_n(r')-\varphi_n(r)]} \ .
  \label{ortho_eigen2}
\eeqa
We now choose the input states for Alice's and Bob's particles to
be
\beq
    \ket{\psi}_A \ket{\phi_r}_B
    = \frac{1}{\sqrt{N}}\sum_{n=1}^N e^{i\eta_n} \ket{n}_A\ket{\lambda_r}_B
    \label{input2}
\eeq
for arbitrary, real parameters $\eta_n$.  Using \eqr{diag} we
find that the output state of the gate will then be
\beqa
  \frac{1}{\sqrt{N}}\sum_{n=1}^N e^{i\eta_n}\ket{n}_A U_n \ket{\lambda_r}_B
  &=& \frac{1}{\sqrt{N}}\sum_{n=1}^N e^{i\eta_n}
        \ket{n}_A e^{i\varphi_n(r)} U_1\ket{\lambda_r}_B \non \\
  &=& \ket{\Psi_r}_A \ket{\Phi_r}_B  \label{output2}
\eeqa
where
\beqan
  \ket{\Psi_r}_A &\equiv& \frac{1}{\sqrt{N}}\sum_{n=1}^N
             e^{i[\varphi_n(r)+\eta_n]}\ket{n}_A    \\
  \ket{\Phi_r}_B &\equiv& U_1\ket{\lambda_r}_B\ .
\eeqan
The output states of Alice's particle $\ket{\Psi_r}_A$ are
orthogonal over $r$ according to \eq{\ref{ortho_eigen2}},
\beq
 {}_A\!\ip{\Psi_r|\Psi_{r'}}_A
     = \frac{1}{N}\sum_{n=1}^N e^{i[\varphi_n(r')-\varphi_n(r)]}
     = \delta_{r,r'}\ .  \label{orthogA}
\eeq
Hence, provided \eqr{commute} is satisfied, Bob can generate
$N$ different orthogonal states of Alice's particle.  This
completes the proof of sufficiency.

The {\em necessity} of the commuting condition is proved by
beginning with the most general input state for Alice's particle,
allowing Bob's particle to be in one of $N$ orthogonal input
states and then showing that for Alice's particle to end up in
one of $N$ orthogonal output states necessarily requires the
commuting property \eqr{commute}. Let the input state be given by
\beq
   \ket{\psi}_A  \ket{\phi_r}_B
      = \sum_{n=1}^N a_n \ket{n}_A \sum_{m=1}^N b_m(r) \ket{m}_B
\label{2out} \eeq where $\sum_n|a_n|^2=\sum_n|b_n(r)|^2=1$, and
where the parameter $r$ indicates that Bob chooses from some set
of orthogonal states $\{\ket{\phi_r}_B \}$. In order that Alice be
able to resolve the final state of her particle into different
states indexed by $r$, without error, the output state of the gate
must factorise into a product state. The combined output state
${\cal G}(\ket{\psi}_A \ket{\phi_r}_B)$ from \eqr{input1} is
therefore of the form \beqa
  \sum_{n=1}^N a_n \ket{n}_A U_n \ket{\phi_r}_B
    = \ket{\Psi_r}_A \ket{\Phi_r}_B
  \label{output1}
\eeqa
where $\ket{\Psi_r}_A$ and $\ket{\Phi_r}_B$ are normalised
states. The left-hand side factorises only if
\beq
   U_n \ket{\phi_r}_B = \beta_n(r) e^{i\xi_n(r)}  \ket{\Phi_r}_B
   \label{p2P}
\eeq for all $r$, where $\beta_n(r)>0$ and $\xi_n(r)$ are real
parameters. Because of the unitarity of $U_n$ this implies
\beqan
  {}_B\ip{\Phi_r|\Phi_{r'}}_B \beta_n(r)\beta_n(r')
               e^{i[\xi_n(r')-\xi_n(r)]}
      = {}_B\ip{\phi_r|\phi_{r'}}_B
      = \delta_{r,r'}
\eeqan
and since ${}_B\ip{\Phi_r|\Phi_{r}}_B=1$ then $\beta_n(r)=1$. Both
sets of input and output states $\{\ket{\Phi_r}_B\}$ and
$\{\ket{\phi_r}_B\}$ therefore form complete orthonormal sets in
Bob's state space and consequently there exists a fixed unitary
mapping $T'$ such that for all $r=1,...,N$
\[
    \ket{\Phi_r}_B = T' \ket{\phi_r}_B \ .
\]
Together with  $\beta_n(r)=1$ and \eq{\ref{p2P}} this leads to
\beqa
  U_n = \sum_{r=1}^N e^{i\xi_n(r)} \ket{\Phi_r}_{B\ B}\!\bra{\phi_r}
      = T' \sum_{r=1}^N e^{i\xi_n(r)} \ket{\phi_r}_{B\ B}\!\bra{\phi_r} \ ,
  \label{2eigen2}
\eeqa
from which the commuting properties \eqr{commute} and
\eqr{equiv_commute} follow. These commuting properties are
therefore both sufficient and necessary for the symmetric
operation. This completes the proof of Theorems $1$ and $1'$.

{\bf Proof of Theorem 2.}  \eqr{2eigen2} shows that Bob's input
state $\ket{\phi_r}_B$ is necessarily an eigenstate of the
pairwise products $U_n^\dagger U_m$ and so necessarily:
\beq
   \ket{\phi_r}_B = \ket{\lambda_r}_B   \label{eigen_r}
\eeq
where $\ket{\lambda_r}_B$ are defined in \eqr{eigen}.
Substituting this into \eqr{output1} and using \eqr{diag} shows
that the output state of Alice's particle is necessarily
\[
   \ket{\Psi_r}_A = \sum_{n=1}^N a_n e^{i\varphi_n(r)} \ket{n}_A
\]
up to an arbitrary phase factor. Alice must be able to
distinguish $N$ different output states indexed by $r$ and so a
necessary condition is
\[
   {}_A\ip{\Psi_{r'} | \Psi_r}_A
     = \sum_{n=1}^N |a_n|^2 e^{i[\varphi_n(r)-\varphi_n(r')]}
     = \delta_{r,r'} \ .
\]
Multiplying by $e^{-i\varphi_m(r)}$, summing over $r$ and using
\eqr{ortho_eigen1} yields
\[
   |a_m|^2 N e^{-i\varphi_m(r')} = e^{-i\varphi_m(r')}
\]
i.e. $a_m=e^{i\eta'_n}/\sqrt{N}$ where  the $\eta'_n$ are
arbitrary, real parameters, and so a necessary condition for
Alice's input state is that \beq
    \ket{\psi}_A
      = \frac{1}{\sqrt{N}} \sum_{n=1}^N e^{i\eta'_n}\ket{n}_A
      \ .  \label{stateA}
\eeq

Eqs. (\ref{input2},\ref{output2},\ref{orthogA}) show that
\eqr{eigen_r} and \eqr{stateA} are also sufficient for Bob to
send one of $N$ distinct messages to Alice.  Hence Theorem 2 is
proved.

{\bf Proof of Corollary 3.}  Substituting \eqr{Um} into
\eqr{ortho1} gives
\beq
   U_n\ket{R}_B = U_1 \sum_{r=1}^N e^{i\varphi_n(r)}
       {}_B\!\ip{\lambda_r|R}_B \ket{\lambda_r}_{B} = \ket{n}_B
       \label{state_n}
\eeq
and hence, from the orthogonality of the sets $\{\ket{n}_B\}$ and
$\{\ket{\lambda_r}_B\}$,
\[
   \delta_{n,m} = \sum_{r=1}^N e^{i[\varphi_n(r)-\varphi_{m}(r)]}
       |{}_B\!\ip{\lambda_r|R}_B|^2 \ .
\]
Multiplying by $e^{-i\varphi_n(r')}$, summing over $n$ and using
\eqr{ortho_eigen2} then gives
\[
   e^{-i\varphi_m(r')} = N e^{-i\varphi_m(r')} |{}_B\!\ip{\lambda_{r'}|R}_B|^2
\]
and so
\beq
  {}_B\!\ip{\lambda_r|R}_B=\frac{1}{\sqrt{N}}e^{-i\gamma_r}
  \label{overlapR}
\eeq
for arbitrary, real $\gamma_r$, which proves \eqr{Rstate}.
\eqr{exp_setN} then follows from \eqr{state_n} and
\eqr{overlapR}. Substituting \eqr{exp_setN} into the right-hand
side of \eqr{lambda} and then performing the sum over $n$ using
\eqr{ortho_eigen2} verifies the equality in \eqr{lambda}.

\section{Discussion and Conclusion} \label{discussion}

Up to now we have concentrated on symmetric orthogonal gates. We
now discuss a result that applies to all orthogonal gates
including those that are {\em asymmetric}. As before let Alice
have the control of an orthogonal gate of cardinality $N$. Imagine
a scenario where $N_B$ is the number of classical messages (not
necessarily the maximum) that Bob can choose to send to Alice and
let the set of the $N_B$ orthogonal input states that Bob uses for
this be $\Lambda\equiv\{\ket{\phi_r}_B:r=1,\ldots,N_B\}$. Further,
let Bob choose to send the $r$th message and let Alice's input
state be given, as before, by the general state $\sum_{n}
a_n\ket{n}_A$. For Alice to be able to distinguish this message
from all others, the output of the gate must factorise into the
form given by \eqr{output1}. \eqr{p2P} then follows necessarily
for $n\in {\cal R}$ where ${\cal R}\equiv\{n:a_n\ne 0\}$. This
implies that \beq
  U^\dagger_m U_n \ket{\phi_r}_B
  = e^{i[\xi_n(r)-\xi_m(r)]}\ket{\phi_r}_B
  \label{assym}
\eeq for $n,m\in {\cal R}$ and $r=1,\ldots,N_B$. The cardinality
of the set ${\cal R}$ gives an upper bound on the dimension of the
subspace in which the final state $\sum_n a_n
e^{i\xi_n(r)}\ket{n}_A$ of Alice's particle lies, and hence must
be at least $N_B$ for Alice to be able to distinguish this many
messages. We note that the operator products on the left-hand side
of \eqr{assym} could share more eigenstates than the $N_B$ states
in $\Lambda$.  Thus a {\em necessary} condition for Bob to send
$N_B$ messages to Alice is that the set ${\cal R}$ must contain at
least $N_B$ elements and all the elements in $\{U^\dagger_m
U_n:n,m\in{\cal R}\}$ must share at least $N_B$ eigenstates.

We have already shown that a controlled-Pauli gate is asymmetric
and that Bob cannot send 2 bits of classical information to Alice
per application of the gate if Alice has the control qubits. Armed
with this necessary condition we can now determine just how much
classical information Bob can transmit to Alice.  For Bob to be
able to send 1 of 3 distinct messages, the set ${\cal R}$ must
contain at least 3 elements and all the elements in $\{U^\dagger_m
U_n:n,m\in{\cal R}\}$ must share at least 3 eigenstates. It is
easy to show that this necessary condition is not fulfilled for
the operators
$U_n\in\{\id_1\otimes\id_2,\sigma_x\otimes\id_2,\sigma_y\otimes\id_2,
\sigma_z\otimes\id_2\}$ of the controlled-Pauli gate.  Hence Bob
cannot send 1 choice from 3 distinct messages.  In the
Introduction we gave a protocol that allows Bob to send 1 choice
from 2 distinct messages to Alice. This is therefore the {\em
maximum} he can send. Thus Bob is limited to sending only one bit
of classical information back to Alice as opposed to the two bits
she could send.

In conclusion, we have defined a class of controlled quantum
gates, which we call orthogonal gates, obeying the property
\eqr{ortho1}. We investigated the algebraic structure that makes
these gates symmetric with respect to the interchange of the
control and target systems. We have shown that all such gates obey
the commuting properties \eqr{commute} and \eqr{equiv_commute} and
that these properties are sufficient. We also discussed a
necessary condition that is useful for determining the maximum
number of classical messages able to be sent using asymmetric
orthogonal gates. We hope that this initial research may shed
light onto the more general problem of establishing the classical
communication capacity of general, not necessarily controlled,
multiparty quantum gates.

{\bf Acknowledgements.} The authors thank A. Chefles, P.
Papadopoulos and M.B. Plenio for discussions. This work has been
supported by the Engineering and Physical Sciences Research
Council (EPSRC) and DGICYT Project No. PB-98-0191 (Spain).\\


\begin{thebibliography}{99}
\bibitem{gates} A. Barenco et al, Phys. Rev. A{\bf 52}, 3457 (1995).
%
\bibitem{distri} J.I. Cirac, A.K. Ekert, S.F. Huelga and C. Macchiavello, Phys.
Rev. A{\bf 59}, 4249 (1999).
%
\bibitem{innsbruck} J.I.~Cirac et al, Phys. Rev. Lett. {\bf 78},
3221 (1997). S.J. van Enk, J.I. Cirac and P. Zoller, Science {\bf
279}, 205 (1998). H.J.~Briegel et al. Phys. Rev. Lett. {\bf 26},
5932 (1998).
%
\bibitem{tony} A. Chefles, C. R. Gilson and S. M. Barnett, 'Entanglement, Information and Multiparticle Quantum Operations', quant-ph/0003062.
%
\bibitem{poly} J. Eisert,
K. Jacobs, P. Papadopoulos and M.B. Plenio, Phys. Rev. A{\bf 62}, 052317
(2000).
%
\bibitem{popescu} D. Collins, N. Linden and S. Popescu, 'The non-local content
of  quantum operations', quant-ph/0005102.
%
\bibitem{cirac} J.~I. Cirac, W. D\"{u}r, B. Kraus and M. Lewenstein,
'Entangling operations and their implementation using a small amount of
entanglement', quant-ph/0007057.
%
\bibitem{mbp} M.B. Plenio invented this gate for the discussion of the minimal
resources required for the remote implementation of an arbitrary
unitary operation. (See S.F. Huelga et al, quant-ph/0005061).
%
\bibitem{superdense}
C.H.~Bennett and S.J.~Wieser, Phys. Rev. Lett. {\bf 69}, 2881,
(1992).
%
\bibitem{note_U_1}
The analysis is easily extended to the general products
$U_n^\dagger U_m$ for an arbitrary value of $n$, but, for
clarity, we treat $n=1$ here.

\end{thebibliography}
\end{document}